\documentclass[conference]{IEEEtran}
\IEEEoverridecommandlockouts

\usepackage{cite}
\usepackage{amsmath,amssymb,amsfonts}
\usepackage{algorithm}
\usepackage{algorithmic}
\usepackage[hidelinks]{hyperref}

\usepackage{multirow}
\usepackage{booktabs}

\usepackage{graphicx}
\usepackage{textcomp}
\usepackage{xcolor}

\def\BibTeX{{\rm B\kern-.05em{\sc i\kern-.025em b}\kern-.08em
    T\kern-.1667em\lower.7ex\hbox{E}\kern-.125emX}}
\begin{document}

\title{SABR: A Stable Adaptive Bitrate Framework \\
Using Behavior Cloning Pretraining and Reinforcement Learning Fine-Tuning\\
\thanks{Code: \url{https://github.com/luopeng69131/SABR} \par
Dataset: \url{https://github.com/luopeng69131/ABRBench}}
}

\author{
    \IEEEauthorblockN{Pengcheng Luo\IEEEauthorrefmark{1}\IEEEauthorrefmark{2}, 
                     Yunyang Zhao\IEEEauthorrefmark{1}\IEEEauthorrefmark{2}, Bowen Zhang\IEEEauthorrefmark{1}\IEEEauthorrefmark{2}, 
                     Genke Yang\IEEEauthorrefmark{1}\IEEEauthorrefmark{2}, \\
                     Boon-Hee Soong\IEEEauthorrefmark{3}, Senior Member, IEEE,
                     Chau Yuen\IEEEauthorrefmark{3}, Fellow, IEEE}
    \IEEEauthorblockA{\IEEEauthorrefmark{1}Ningbo Artificial Intelligence Institute, Shanghai Jiao Tong University, Ningbo, China \\
    \IEEEauthorblockA{\IEEEauthorrefmark{2}School of Automation and Intelligent Sensing, Shanghai Jiao Tong University, Shanghai, China\\
    Email: \{luopeng69131, zyyfighting, bwz96sco, gkyang\}@sjtu.edu.cn}}
    \IEEEauthorblockA{\IEEEauthorrefmark{3}School of Electrical and Electronic Engineering, Nanyang Technological University, Singapore \\
    Email: \{ebhsoong, chau.yuen\}@ntu.edu.sg}
}

\maketitle

\begin{abstract}
With the advent of 5G, the internet has entered a new video-centric era. From short-video platforms like TikTok to long-video platforms like Bilibili, online video services are reshaping user consumption habits. Adaptive Bitrate (ABR) control is widely recognized as a critical factor influencing Quality of Experience (QoE). Recent learning-based ABR methods have attracted increasing attention. However, most of them rely on limited network trace sets during training and overlook the wide-distribution characteristics of real-world network conditions, resulting in poor generalization in out-of-distribution (OOD) scenarios. To address this limitation, we propose SABR, a training framework that combines behavior cloning (BC) pretraining with reinforcement learning (RL) fine-tuning. We also introduce benchmarks, ABRBench-3G and ABRBench-4G+, which provide wide-coverage training traces and dedicated OOD test sets for assessing robustness to unseen network conditions. Experimental results demonstrate that SABR achieves the best average rank compared with Pensieve, Comyco, and NetLLM across the proposed benchmarks. These results indicate that SABR enables more stable learning across wide distributions and improves generalization to unseen network conditions.

\end{abstract}


\section{Introduction}

\begin{figure}[htbp]
\includegraphics[width=1\linewidth]{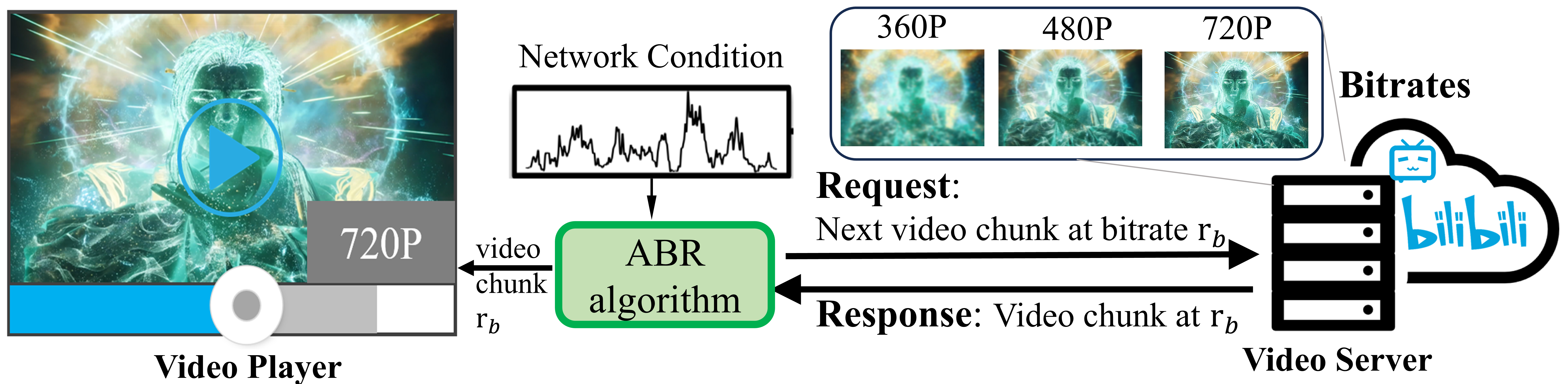}
\caption{An overview of ABR.}
\label{fig-abr}
\end{figure}

The emergence of 5G networks marks a new stage of internet development, in which video constitutes the dominant share of digital content. Short-form services such as TikTok and long-form streaming platforms such as Bilibili are reshaping content consumption habits, making video the primary medium for information, entertainment, and social interaction worldwide. In this context, the smoothness and clarity of video playback are decisive for user experience, with Adaptive Bitrate (ABR) algorithms serving as a fundamental mechanism to ensure high Quality of Experience (QoE). ABR algorithms dynamically adjust video bitrate in response to real-time fluctuations in network bandwidth, thereby minimizing stalling and latency, as illustrated in Figure~\ref{fig-abr}.

As the user base continues to expand, video streaming service providers accumulate massive volumes of network data on a daily basis. This wealth of data presents unprecedented opportunities for analyzing user behavior and optimizing streaming strategies, while also providing a solid foundation for applying artificial intelligence (AI) techniques to ABR research. AI approaches such as deep learning and reinforcement learning (RL) are increasingly driving ABR algorithms toward higher performance and stronger adaptability. Nevertheless, current research still faces the following two major challenges:
\begin{itemize}
    \item Limited generalization to unseen distributions: Most studies train ABR models on a specific network trace set, without fully leveraging the vast amount of network trace data. Therefore, models exhibit limited performance when facing unseen network conditions.
    \item Degradation under wide-distribution training: When the training dataset encompasses a broad spectrum of network conditions, the efficiency and stability of the ABR model training can be significantly undermined.
\end{itemize}

Similar issues have been studied in the field of large language models (LLMs), where the two-stage training framework of pretraining + fine-tuning has proven to be an effective solution~\cite{radford2018improving,devlin2019bert}. The pretraining stage enables the model to acquire initial representations and understanding of wide-distribution training data, while the fine-tuning stage enables more effective generalization to the target environment. In LLM alignment techniques, Supervised Fine-Tuning (SFT) + Reinforcement Learning from Human Feedback (RLHF) can be regarded as an extension of this framework~\cite{ouyang2022training}. SFT uses large-scale supervised data to help the model initially understand human instructions and task structures, while RLHF leverages the exploration capability of RL to align the model’s behavior with human preferences. This combination enables Generative Pre-trained Transformer (GPT) models to faithfully assist and serve humans in real-world daily applications.

Inspired by this, we propose a two-stage training framework for ABR, termed SABR: Behavior Cloning (BC) pretraining + RL fine-tuning. In the pretraining stage, we adopt the Direct Preference Optimization (DPO)~\cite{rafailov2023direct} algorithm to perform BC on expert data, obtaining a base model. In the fine-tuning stage, we optimize the base model using the Proximal Policy Optimization (PPO)~\cite{schulman2017proximal} algorithm. We also integrate mainstream network trace sets and videos to construct benchmarks: ABRBench-3G and ABRBench-4G+. Each benchmark contains a training set, a test set, and an Out-of-Distribution (OOD) set. Our main contributions are as follows:
\begin{itemize}
    \item We propose a stable framework, SABR, which combines BC pretraining and RL fine-tuning. The framework improves ABR generalization by leveraging a wide range of network trace data. 
    
    \item We design SABR with DPO-based BC for fast and stable pretraining, and PPO-based RL for deeper exploration, enabling robust adaptation to challenging network dynamics.

    \item We release two benchmarks, which provide an effective evaluation of ABR models’ generalization to unseen network conditions.

    \item We empirically validate that SABR achieves the best average rank compared with Pensieve, Comyco, NetLLM, and the other baselines.
    

\end{itemize}


\section{Related works}
Learning-based ABR research has been extensively explored, with the core idea of leveraging neural networks and RL to overcome the limitations of traditional rule-based bitrate control. Pensieve~\cite{mao2017neural} was the first to apply the RL model to ABR, using network states (e.g., throughput and buffer length) as inputs to train an A3C~\cite{mnih2016asynchronous} policy on 3G network traces, thereby demonstrating the feasibility and advantages of RL in ABR control. Comyco~\cite{huang2019comyco} further introduced quality-aware QoE metrics and employed imitation learning from Model Predictive Control (MPC)-generated expert data, significantly improving training efficiency and model performance. To address user differences in video quality preferences, Jade~\cite{huang2023optimizing} incorporated ranking-based QoE feedback into RLHF, aligning the optimization objective and achieving QoE improvements across heterogeneous network conditions. 


Genet~\cite{xia2022genet} introduced an automatic curriculum learning approach~\cite{bengio2009curriculum}, which starts from network environments with large performance gaps compared to the rule baselines, and gradually expands the training distribution, thereby enabling the model to improve progressively. However, curriculum learning may suffer from distributional shift and forgetting issues when the training distribution becomes broad. NetLLM~\cite{wu2024netllm} adapted LLMs to multiple networking tasks, including ABR. Through multi-modal encoding and Low-Rank adaptation (LoRA)~\cite{hu2022lora}, it reduced training costs and showcased the potential of LLMs in ABR tasks. 


While these works have advanced learning-based ABR, two limitations persist: limited generalization to unseen network conditions and degraded stability under wide-distribution training. These issues underscore the necessity of more robust and efficient training paradigms, with comprehensive benchmarks for evaluation.

\section{Proposed SABR framework}
\begin{figure}[htbp]
\includegraphics[width=1\linewidth]{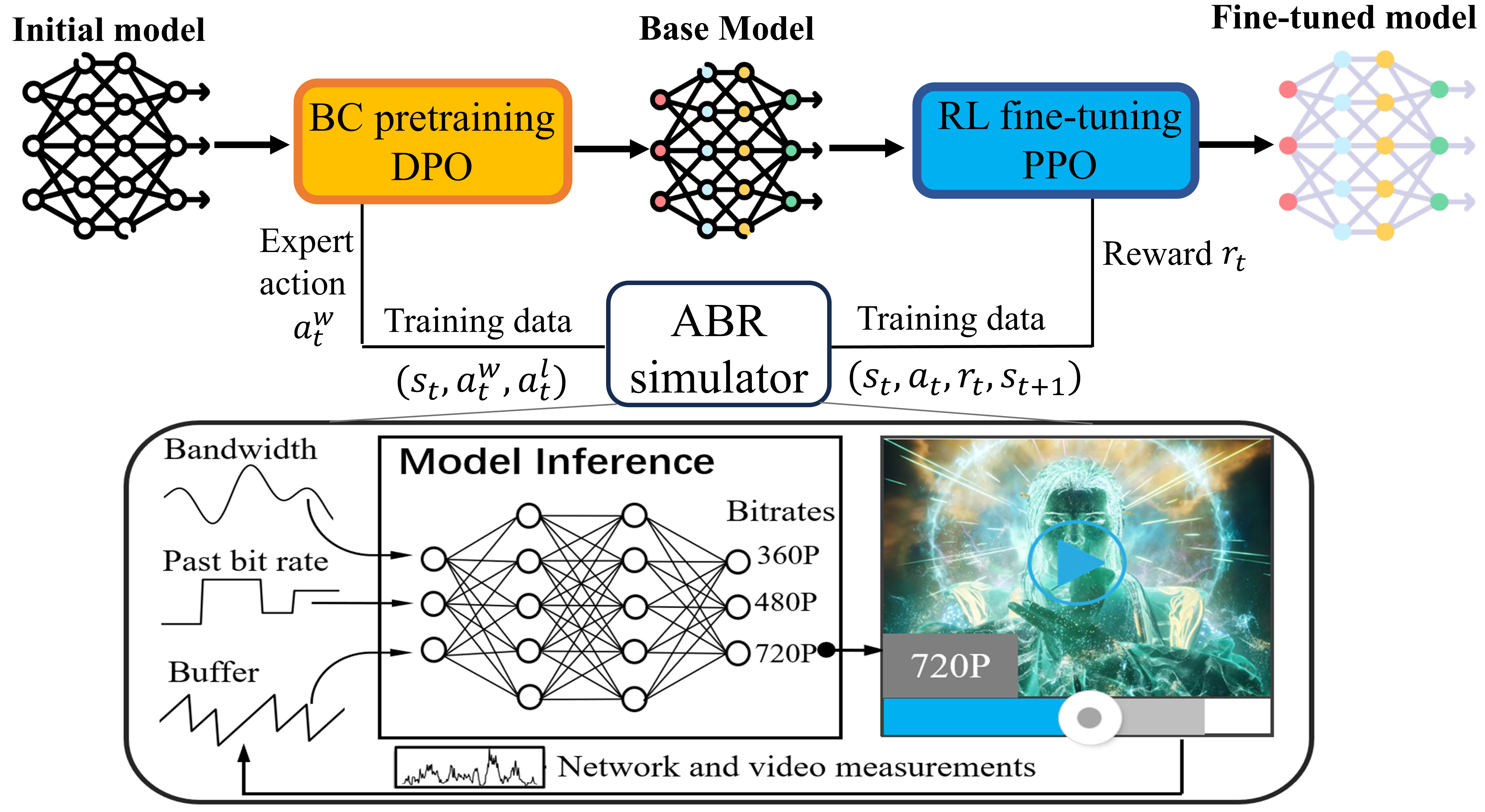}
\caption{Proposed SABR framework: BC pretraining + RL fine-tuning.}
\label{fig-sabr}
\end{figure}

The SABR framework consists of two stages: BC pretraining and RL fine-tuning. In the BC pretraining stage, we train the model on expert data using the DPO algorithm to obtain a base model. In the RL fine-tuning stage, we refine the base model via PPO training. An overview of the framework is shown in Figure~\ref{fig-sabr}.

\subsection{BC pretraining with DPO}


Originally proposed for preference alignment in LLMs, DPO directly maximizes the likelihood ratio of human-preferred responses, thereby avoiding the need for reward models and complex RL optimization commonly used in traditional RLHF pipelines. Motivated by its ability to directly capture preferences from data, we adopt DPO to learn from expert demonstrations for ABR. In the BC pretraining stage, we use DPO to efficiently learn from expert samples, treating them as preferred actions. This initializes a base model with stable performance and a stronger control policy for ABR.

In the original DPO algorithm, given a pair of candidate trajectories $\tau_w$ (the “winner”) and $\tau_l$ (the “loser”), it directly maximizes the log-ratio of their probabilities to favor the preferred trajectory. The objective is defined as:
\begin{align}
\mathcal{L}_{\text{DPO}}(\theta) = 
- \mathbb{E}_{(\tau_w, \tau_l) \sim \mathcal{D}} \Big[ 
\log \sigma \Big( \beta \cdot \Big[ &
\log \frac{ \pi_\theta(\tau_w) }{ \pi_{\text{ref}}(\tau_w) } \notag \\
& - 
\log \frac{ \pi_\theta(\tau_l) }{ \pi_{\text{ref}}(\tau_l) } 
\Big] \Big) \Big].
\end{align}
Here, $\pi_\theta(\tau)$ denotes the likelihood of trajectory $\tau$ under the current model, while $\pi_{\text{ref}}(\tau)$ represents the likelihood under a reference model, typically the initialization model. The scalar $\beta > 0$ controls the update strength, and $\sigma(\cdot)$ denotes the sigmoid function. $\mathcal{D}$ is the set of preference trajectory pairs.

In BC training, since we focus on learning from each state-action pair, we adapt the original DPO loss into a step-wise formulation as follows:
\begin{align}
\mathcal{L}_{\text{DPO-step}}(\theta) = 
- \mathbb{E}_{(s, a^w, a^l) \sim \mathcal{D}} \Big[ 
\log \sigma \Big( 
\beta \cdot \Big[ &
\log \frac{\pi_\theta(a^w \mid s)}{\pi_{\text{ref}}(a^w \mid s)} \notag \\
& - 
\log \frac{\pi_\theta(a^l \mid s)}{\pi_{\text{ref}}(a^l \mid s)} 
\Big] 
\Big) 
\Big].
\label{equ-dpo_step}
\end{align}
Here, $(s, a^w, a^l) \sim \mathcal{D}$ are sampled state-action pairs, where $a^w$ is an expert (preferred) action and $a^l$ is a less preferred (e.g., randomly sampled) alternative. The loss encourages the model to increase the preference margin for expert actions over less preferred ones at each step. 

The BC training procedure is designed following the DAGGER algorithm~\cite{ross2011reduction}, as detailed in Algorithm~\ref{alg:BC_training}. Through interaction with the ABR simulator, the model collects samples that are subsequently used for training. The beam search strategy follows the implementation from Comyco~\cite{huang2019comyco,godka2025comyco-lin}.

\begin{algorithm}
\caption{BC pretraining with DPO}
\label{alg:BC_training}
\begin{algorithmic}[1]
\STATE \textbf{Input:} Initial model $\pi_{\theta}$, \textsc{Beam\_Search\_Policy}, ABR simulator, iteration $N_{\text{pretrain}}$, rollout step $T_{\text{pretrain}}$, epoch $E_{pretrain}$, mini-batch size $m_{pretrain}$
\STATE Initialize $\pi_{ref}$, buffer $\mathcal{B} \gets \varnothing$, obtain initial state $s_1$ from ABR simulator
\FOR{$1,2,\dots,N_{\text{pretrain}}$} 
    \FOR{$1,2,\dots,T_{\text{pretrain}}$}
        \STATE Select action $a_t \sim \pi_{\theta}(\cdot \mid s_t)$
        \STATE Expert action $a_t^w \gets$ \textsc{Beam\_Search\_Policy}($s_t$) 
        \STATE Randomly select an alternative action $a_t^l \neq a_t^w$
        \STATE Append sample: $\mathcal{B} \gets \mathcal{B} \cup \{(s_t, a_t^w, a_t^l)\}$
        \STATE Execute $a_t$ in the ABR simulator to obtain next state $s_{t+1}$
    \ENDFOR

    \FOR{$1,2,\dots,E_{pretrain}$}
        \STATE Sample mini-batch $\hat{\mathcal{B}}$ of size $m_{pretrain}$ from $\mathcal{B}$
        \STATE Update $\pi_{\theta}$ using the DPO loss on $\hat{\mathcal{B}}$ (Eq.~\ref{equ-dpo_step})
    \ENDFOR
\ENDFOR
\STATE \textbf{Output:} Base model $\pi_{\theta}$
\end{algorithmic}
\end{algorithm}

\subsection{RL fine-tuning with PPO}
Only BC training is constrained to the distribution of expert policies and lacks the capacity to explore a broader policy space. To improve generalization in network environments, we perform RL fine-tuning of the base model using PPO. PPO is a policy-gradient–based RL method that restricts the extent of policy updates between iterations to prevent training instability and performance collapse. PPO has demonstrated strong stability and sample efficiency in both continuous~\cite{todorov2012mujoco} and discrete tasks~\cite{luo2022multi}.

PPO consists of both an actor network $\pi_\theta$ and a critic network $V_\phi$. The objective of the actor network is formalized through the actor loss, given by:
\begin{equation}
L^{\text{Actor}}(\theta) = 
\mathbb{E}_t \Big[ 
\min \big( r_t(\theta) A_t, \;
\text{clip}(r_t(\theta), 1-\epsilon, 1+\epsilon) A_t \big) 
\Big],
\end{equation}
where 
\begin{equation}
r_t(\theta) = 
\frac{\pi_\theta(a_t \mid s_t)}{\pi_{\theta_{\text{old}}}(a_t \mid s_t)},
\end{equation}
denotes the probability ratio between the current actor network $\pi_\theta(a \mid s)$ and the previous actor network $\pi_{\theta_{\text{old}}}(a \mid s)$. $A_t$ is the advantage estimate at time step $t$, and $\epsilon$ is the clipping threshold. The advantage function $A_t$ is typically computed using Generalized Advantage Estimation (GAE)~\cite{schulman2015high}, which reflects the reward information that guides policy improvement. 

\begin{algorithm}
\caption{RL fine-tuning with PPO}
\label{alg:ppo_finetune}
\begin{algorithmic}[1]
\STATE \textbf{Input:} Actor network $\pi_{\theta}$ (initialized from base model), critic network $V_{\phi}$, ABR simulator, iteration $N_{\text{finetune}}$, rollout steps $T_{\text{finetune}}$, PPO epochs $E_{\text{finetune}}$, mini-batch size $m_{\text{finetune}}$, clipping parameter $\epsilon$, discount factor $\gamma$, GAE parameter $\lambda$
\STATE Empty buffer $\mathcal{B} \gets \varnothing$, obtain initial state $s_1$ from ABR simulator
\FOR{$1,2,\dots,N_{\text{finetune}}$}
    \FOR{$1,2,\dots,T_{\text{finetune}}$}
        \STATE Select action $a_t \sim \pi_{\theta}(\cdot \mid s_t)$
        \STATE Execute $a_t$ in the ABR simulator to obtain reward $r_t$ and next state $s_{t+1}$
        \STATE Append transition: $\mathcal{B} \gets \mathcal{B} \cup \{(s_t, a_t, r_t, s_{t+1})\}$
    \ENDFOR


    \STATE For all transitions in $\mathcal{B}$, compute $\hat V_t = V_{\phi}(s_t)$ and $\hat V_{t+1} = V_{\phi}(s_{t+1})$
    \STATE Compute TD errors $\delta_t = r_t + \gamma \hat V_{t+1} - \hat V_t$, then advantages $\hat A_t$ via GAE with $(\gamma,\lambda)$
    \STATE Set target value $V_t^{\text{target}} = \hat V_t + \hat A_t$ for critic updates

    \STATE Augment each transition in $\mathcal{B}$ to $\{(s_t, a_t, r_t, s_{t+1}, \hat{A}_t, V_t^{\text{target}})\}$

    \FOR{$1,2,\dots,E_{\text{finetune}}$}
        \STATE Sample mini-batch $\hat{\mathcal{B}}$ of size $m_{\text{finetune}}$ from $\mathcal{B}$
        \STATE Update parameters $\theta$ and $\phi$ using the full PPO objective (Eq.~\ref{equ-ppo-loss}) on $\hat{\mathcal{B}}$
    \ENDFOR

    \STATE Clear $\mathcal{B} \gets \varnothing$
    \STATE $\pi_{\theta_{\text{old}}} \gets \pi_{\theta}$
\ENDFOR
\STATE \textbf{Output:} fine-tuned model $\pi_{\theta}$
\end{algorithmic}
\end{algorithm}

The full PPO objective combines the actor loss, critic loss, and an entropy regularization term, and is given by:
\begin{equation}
L^{\text{PPO}}(\theta) = 
\mathbb{E}_t \Big[
L^{\text{Actor}}(\theta) 
- c_1 \big( V_\phi(s_t) - V_t^{\text{target}} \big)^2
+ c_2 S[\pi_\theta](s_t)
\Big],
\label{equ-ppo-loss}
\end{equation}
where $V_\phi(s_t)$ is the state value predicted by the critic network, with $\big( V_\phi(s_t) - V_t^{\text{target}} \big)^2$ as the critic loss where $V_t^{\text{target}}$ is the target value; $S[\pi_\theta](s_t)$ is an entropy regularization term encouraging exploration; and $c_1$ and $c_2$ are their respective weighting coefficients. The overall RL fine-tuning procedure with PPO is shown in Algorithm~\ref{alg:ppo_finetune}.

\section{Proposed benchmarks}
We release two benchmarks: ABRBench-3G and ABRBench-4G+. Each benchmark consists of both video content and network traces. The traces are reorganized and curated from publicly available trace sets on the internet, such as Lumos 4G/5G~\cite{narayanan2021variegated,narayanan2020lumos5g} and FCC~\cite{riiser2013commute,mao2017neural,FCC2016Broadband}. Each benchmark contains multiple trace sets to ensure broad coverage of network conditions.

In each benchmark, traces are divided into training, testing, and OOD sets. The training and testing sets are created by splitting each trace set proportionally. For example, in FCC-18, 75\% of traces are allocated to the training set, while the remaining 30\% are used for testing. The OOD set is also used to evaluate model performance, but unlike the test set, it specifically focuses on assessing generalization to unseen distributions. Therefore, trace sets included in the OOD set are not split or reused in other sets.

For training, models are trained on the entire training set with all traces randomly shuffled. Evaluation is performed separately for each trace set within the test and OOD sets. During evaluation, we preserve the trace set granularity, since certain trace sets (e.g., those with high bandwidth) can skew the overall average QoE and mask the performance under other bandwidth conditions. Tables~\ref{tab-abr3g} and \ref{tab-abr4g} present the trace set information of ABRBench-3G and ABRBench-4G+.


\begin{table}[htbp]
\caption{ABRBench-3G trace statistics}
\label{tab-abr3g}
\begin{center}
\begin{tabular}{l|lcl}
\toprule
\textbf{Group} & \textbf{Trace Set}   & \textbf{Count} & \textbf{Range (Mbps)} \\
\midrule
Training       & Same with test & 1828             & $0.00 \sim 45.38$              \\
\midrule
\multirow{5}{*}{Test} & FCC-16~\cite{riiser2013commute,mao2017neural,FCC2016Broadband}         & 69             & $0.00 \sim 8.95$                 \\
               & FCC-18~\cite{FCC2018RawDataReleases,meng2019pitree}       & 100             & $0.00 \sim 41.76$                 \\
               & Oboe~\cite{akhtar2018oboe,kan2022improving}          & 100             & $0.16 \sim 9.01$                 \\
               & Puffer-21~\cite{yan2020learning,kan2022improving}      & 100             & $0.00 \sim 25.14$                 \\
               & Puffer-22~\cite{yan2020learning,kan2022improving}   & 100             & $0.00 \sim 9.29$                 \\
\midrule
OOD            & HSR ~\cite{meng2019pitree}           & 34             & $0.00 \sim 44.68$                 \\

\bottomrule
\end{tabular}
\end{center}
\end{table}

\begin{table}[htbp]
\caption{ABRBench-4G+ trace statistics}
\label{tab-abr4g}
\begin{center}
\begin{tabular}{l|lcl}
\toprule
\textbf{Group} & \textbf{Trace Set}   & \textbf{Count} & \textbf{Range (Mbps)} \\
\midrule
Training       & Same with test & 262             & $0.00 \sim 1890.00$ \\
\midrule
\multirow{3}{*}{Test} & Lumos 4G~\cite{narayanan2021variegated,narayanan2020lumos5g}  & 53 & $0.00 \sim 270.00$  \\
                      & Lumos 5G~\cite{narayanan2021variegated,narayanan2020lumos5g}  & 37 & $0.00 \sim 1920.00$ \\
                      & Solis Wi-Fi~\cite{lv2022lumos}                               & 24 & $0.00 \sim 124.00$  \\
\midrule
\multirow{2}{*}{OOD}  & Ghent~\cite{meng2019pitree}                                    & 40 & $0.00 \sim 110.97$  \\
                      & Lab~\cite{meng2019pitree}                                  & 61 & $0.16 \sim 175.91$  \\

\bottomrule
\end{tabular}
\end{center}
\end{table}

We denote the set of available bitrates as $R$. Specifically, ABRBench-3G uses the Envivio-Dash3~\cite{dashjs} video with $R^{3G} = \{300, 750, 1200, 1850, 2850, 4300\}$, while ABRBench-4G+ uses the Big Buck Bunny~\cite{bigbuckbunny2008} video with $R^{4G+} = \{1000, 2500, 5000, 8000, 16000, 40000\}$.

\section{Implementation details}
The state, action, reward function, and state transition in our Markov Decision Process are consistent with those in Pensieve~\cite{mao2017neural}. Our ABR simulator follows the design of Pensieve’s Python environment~\cite{mao2017neural}, while using the C++ implementation from~\cite{huang2019comyco,godka2025comyco-lin} to improve efficiency. Apart from the C++ simulator, all other components are implemented in Python.

The BC pretraining is implemented in PyTorch~\cite{paszke2019pytorch}, while RL fine-tuning is based on the PPO algorithm from Stable-Baselines3 (SB3)~\cite{raffin2021stable}. During training, we utilize the Vector Environment module of SB3 to enable parallel sample collection, thereby improving training efficiency. The number of parallel environments is set to 4.

In the implementations of Pensieve~\cite{mao2017neural} and Comyco~\cite{huang2019comyco}, the input features are represented as a 6-by-8 matrix. In our implementation, we flatten this matrix into a 48-dimensional vector. The actor network $\pi_\theta$ (base model) adopts a fully connected network of $[48, \tanh, 64, \tanh, 64, 6]$, while the critic network is designed as $[48, \tanh, 64, \tanh, 64, 1]$. The two networks do not share parameters. For both DPO and PPO training, the Adam optimizer~\cite{kingma2014adam} is employed. The hyperparameter settings of the SABR are shown in Table~\ref{tab:hparams}.

\begin{table}[t]
\caption{Hyperparameters for the SABR framework}
\label{tab:hparams}
\centering
\begin{tabular}{llc}
\toprule
\textbf{Symbol} & \textbf{Description} & \textbf{Value} \\
\hline
\multicolumn{3}{c}{DPO parameters} \\
\hline
$N_{\text{pretrain}}$   & Iteration (DPO)                & 15  \\
$E_{\text{pretrain}}$   & Epochs per pretraining iteration                    & 5   \\
$T_{\text{pretrain}}$   & Rollout steps per iteration             & 2000 \\
$m_{\text{pretrain}}$   & Mini-batch size (pretraining)                & 128  \\
$\alpha_{\text{pretrain}}$                & DPO learning rate                                  & 3e-4 \\
$\beta$                 & DPO update scale                                   & 0.1 \\

\hline
\multicolumn{3}{c}{PPO parameters} \\
\hline
$N_{\text{finetune}}$   & Iteration (PPO)                & 244  \\
$E_{\text{finetune}}$   & PPO epochs per update                               & 10   \\
$T_{\text{finetune}}$   & Rollout steps per environment                       & 512 \\
$m_{\text{finetune}}$   & Mini-batch size (fine-tuning)                 & 64  \\
$\alpha_{\text{finetune}}$                & PPO learning rate                                  & 3e-4 \\
$\epsilon$              & Clipping threshold                                  & 0.2 \\
$\gamma$                & Discount factor                                     & 0.99 \\
$\lambda$               & GAE parameter                                       & 0.95 \\
$c_1$                & Coefficient of critic loss                            & 0.5 \\
$c_2$               & Coefficient of entropy                                 & 0.0 \\
\hline
\multicolumn{3}{c}{Other parameters} \\
\hline
$L_{\text{beam}}$       & Beam search future horizon                     & 5   \\
$K_{\text{max}}$       & Beam search maximum beam                    & 5000   \\

\bottomrule
\end{tabular}
\end{table}

\section{Evaluation}
\subsection{Experimental setup}

We build a trace-driven ABR simulator~\cite{mao2017neural}, where both network traces and video content are drawn from ABRBench-3G and ABRBench-4G+. Each experiment is conducted on videos consisting of 49 chunks, with each chunk lasting 4 seconds, emulated over the collected network traces.

We evaluate performance using the QoE metrics:
\begin{equation}
QoE = \sum_{n=1}^N q(R_n)- \delta \sum_{n=1}^{N-1} \big| q(R_{n+1}) - q(R_n) \big| - \mu \sum_{n=1}^N T_n,
\end{equation}
where $N$ represents the total number of video chunks, $R_n$ is the bitrate of the $n$-th chunk, and $T_n$ denotes the rebuffering time at that step. The function $q(R_n)$ maps the bitrate $R_n$ to a corresponding quality score. $\delta$ is the smoothness penalty coefficient, and $\mu$ is the rebuffering penalty coefficient.


Consistent with prior work~\cite{mao2017neural,huang2019comyco,yin2015control}, we adopt $q(R_n) = R_n$, where $R_n \in R^{3G}$ or $R^{4G+}$. We set $N=49$, $\delta = 1$, and use $\mu = 4.3$ for ABRBench-3G and $\mu = 40$ for ABRBench-4G+. We compare SABR against baselines:  
\begin{itemize}
    \item Buffer-Based (BB): A simple heuristic that adapts bitrates based on buffer occupancy to reduce rebuffering.

    \item BOLA~\cite{spiteri2020bola}: Uses Lyapunov optimization to select bitrates solely considering buffer occupancy observations.
    
    \item RobustMPC~\cite{yin2015control}: An extension of the MPC method. It maximizes a given QoE metric over a horizon of 5 future chunks.

    \item QUETRA~\cite{yadav2017quetra}:  A queueing-theoretic algorithm that models the ABR task as an M/D/1/K system, enabling bitrate decisions based on expected buffer occupancy.

    \item Pensieve~\cite{mao2017neural}: An RL-based ABR method that trains a policy network with A3C to maximize a QoE reward. 

    \item Comyco~\cite{huang2019comyco}: A learning-based ABR method that employs imitation learning to train a policy from MPC-generated expert trajectories.

    \item NetLLM~\cite{wu2024netllm}: Adapts LLMs to ABR by combining parameter-efficient fine-tuning (LoRA) with offline RL.

\end{itemize}
For the comparative evaluation, each algorithm is executed ten times, and the average performance is reported. For the learning-based methods (SABR, Pensieve, Comyco, and NetLLM), each result is obtained by training ten separate models, and the reported performance is the average across all models on the test runs. Furthermore, we compute the average rank of each algorithm across the multiple trace sets in each benchmark. Formally, let $r_{i,j}$ denote the rank of algorithm $i$ on trace set $j$, and let $M$ be the total number of trace sets in the benchmark. The average rank of algorithm $i$ is defined as
\begin{equation}
\text{Ave Rank}(i) = \frac{1}{M} \sum_{j=1}^{M} r_{i,j}.
\end{equation}
A lower average rank indicates better overall performance.

\subsection{Proposed SABR vs. existing baselines}
To evaluate the generalization capability of the models, we conducted comparisons across different methods on the test sets of ABRBench-3G and ABRBench-4G+. The learning-based models were trained on the corresponding benchmark training sets before testing. Tables \ref{tab-3g-test} and \ref{tab-4g-test} show the QoE performance of the different methods.

\begin{table}[h]
\caption{QoE Performance comparison on the ABRBench-3G test sets}
\label{tab-3g-test}
\centering
\resizebox{\linewidth}{!}{%
\begin{tabular}{lccccc|c}
\toprule
\textbf{Algorithm} & \textbf{FCC-16} & \textbf{FCC-18} & \textbf{Oboe} & \textbf{Puffer-21} & \textbf{Puffer-22} & \textbf{Ave Rank} \\ 
\midrule
BB        & 25.37 & 131.54 & 82.74 & -6.05 & 13.28 & 7.2 \\
BOLA      & 32.51 & 123.42 & 81.02 & 38.35 & 30.99 & 6.0 \\
QUETRA    & 33.91 & 122.25 & 82.84 & \textbf{42.48} & 36.89 & 4.4 \\
RobustMPC & 36.56 & 143.30 & 96.14 & 34.13 & 36.90 & 3.4 \\
Pensieve  & 34.50 & 134.39 & 90.92 & 38.94 & 35.23 & 3.8 \\
Comyco    & 32.10 & 143.89 & 96.23 & -4.09 & 31.34 & 4.8 \\
NetLLM    & 21.92 & 141.91 & 97.39 & 37.55 & 33.73 & 4.6 \\
SABR      & \textbf{36.68} & \textbf{145.18} & \textbf{99.68} & 36.05 & \textbf{40.05} & \textbf{1.8} \\
\bottomrule
\end{tabular}
}
\end{table}

\begin{table}[h]
\caption{QoE Performance comparison on the ABRBench-4G+ test sets}
\label{tab-4g-test}
\centering
\begin{tabular}{lccc|c}
\toprule
\textbf{Algorithm} & \textbf{Lumos 4G} & \textbf{Lumos 5G} & \textbf{Solis Wi-Fi} & \textbf{Ave Rank} \\ 
\midrule
BB        & 1255.91 & 1726.66 & 429.34 & 5.0 \\
BOLA      & 1200.05 & 1614.40 & 477.08 & 5.0 \\
QUETRA    &  754.43 &  992.74 & 421.58 & 7.7 \\
RobustMPC & 1283.05 & 1696.77 & \textbf{589.64} & 3.0 \\
Pensieve  & 1160.76 & 1828.24 & 447.84 & 5.0 \\
Comyco    & 1285.43 & \textbf{1835.42} & 552.55 & 2.0 \\
NetLLM    &  672.35 & 1510.35 & 474.15 & 6.7 \\
SABR      & \textbf{1309.65} & 1832.14 & 576.33 & \textbf{1.7} \\
\bottomrule
\end{tabular}
\end{table}

For ABRBench-3G, SABR achieved the best QoE performance on FCC-16, FCC-18, Oboe, and Puffer-22. For ABRBench-4G+, SABR achieved the highest QoE on Lumos 4G, while performing slightly worse than the best methods on the other two trace sets. Across both benchmarks, SABR attains the lowest average rank among all methods, demonstrating its overall superior performance and robustness across diverse network conditions.

\subsection{Evaluation on the OOD datasets}

To evaluate the generalization performance of the models under unseen distributions, we conducted comparisons on the OOD datasets of ABRBench-3G (HSR) and ABRBench-4G+ (Ghent and Lab). The learning-based models were trained on the corresponding benchmark training sets before testing. Table~\ref{tab-ood} presents the QoE performance of the different methods. SABR obtained the lowest average rank (2.0), outperforming Comyco (3.7), RobustMPC (4.0), and other baselines. This indicates that SABR maintains strong performance on unseen distributions.

\begin{table}[h]
\caption{QoE Performance comparison on the OOD sets}
\label{tab-ood}
\centering
\begin{tabular}{lccc|c}
\toprule
\textbf{Algorithm} & \textbf{HSR} & \textbf{Ghent} & \textbf{Lab} & \textbf{Ave Rank} \\ 
\midrule
BB        & 138.86 &  834.30 & 1429.22 & 4.3 \\
BOLA      & 137.02 &  912.39 & 1342.63 & 5.0 \\
QUETRA    & 132.56 &  566.61 &  965.94 & 7.0 \\
RobustMPC & 122.37 & \textbf{1075.17} & 1527.84 & 4.0 \\
Pensieve  & 137.82 &  652.45 & 1508.43 & 4.7 \\
Comyco    & 130.22 &  963.94 & \textbf{1595.09} & 3.7 \\
NetLLM    & 129.25 & 1035.09 & 1307.49 & 5.3 \\
SABR      & \textbf{142.20} & 1023.56 & 1561.18 & \textbf{2.0} \\
\bottomrule
\end{tabular}
\end{table}

\section{Conclusion}
In this paper, we propose SABR, a two-stage framework consisting of BC pretraining and RL fine-tuning. The framework is designed to improve stability and training efficiency under wide-distribution data. In the pretraining stage, we employ DPO to learn from expert demonstrations, which provides the model with an initial understanding of the training distribution and establishes a basic control policy. The fine-tuning stage then applies PPO to further optimize the policy, enhancing generalization to unseen network conditions. We further contribute two benchmarks, ABRBench-3G and ABRBench-4G+, to evaluate performance across wide-distribution data and unseen environments. Experimental results show that, on both benchmarks, SABR achieves the best average rank compared with methods such as Pensieve, Comyco, and NetLLM, demonstrating better generalization performance. In future work, we plan to extend our benchmarks with more traces and videos to provide a more comprehensive evaluation for ABR research.

\bibliographystyle{IEEEtran}
\bibliography{ref}

\vspace{12pt}

\end{document}